\documentstyle[12pt,epsf]{article}
\newcommand{\bm}{\bibitem}

\def\be {\begin{equation}}
\def\ee {\end{equation}}
\def\bea {\begin{eqnarray}}
\def\eea {\end{eqnarray}}
\def\k {|{\vec k}|}
\def\q {|{\vec q}|}

\def\qs {|{{\vec q}|}^\ast}
\def\mqs {M_q^\ast}
\def\eqs {E_q^\ast}

\def\mks {M_k^\ast}
\def\eks {E_k^\ast}

\begin{document}
\normalbaselineskip = 24 true pt
\normalbaselines
\thispagestyle{empty}
\rightline{\large\sf SINP-TNP/98-02}
\rightline{\large Jan 1998}
\vskip 0.5 cm
\begin{center}
{\Large { Pauli coupling of vector meson and softening of the nuclear equation
of state}}
\vskip .6 cm
{\bf Subir Bhattacharyya$^{a}$, Abhee K. Dutt- Mazumder$^{b,}${\footnote
{email: abhee@tnp.saha.ernet.in}},\\
Binayak Dutta-Roy$^b$ and Bikash Sinha$^{b,c}$}\\[.4 cm] 
$^a$ Nuclear Research Laboratory,
Bhabha Atomic Research Centre, Mumbai-400 085, India.
\vskip0.2cm
$^b$Saha Institute of Nuclear Physics\\
1/AF, Bidhan Nagar, Calcutta - 700 064, India\\[0.4 cm]
$^c$Variable Energy Cyclotron Centre\\
1/AF, Bidhan Nagar, Calcutta - 700 064, India\\
\end{center}
\vskip .8 cm
\begin{abstract}
We investigate the equation of state (EOS) for nuclear matter, within
the framework of the Relativistic Hartree Fock (RHF) theory, with
special emphasis on the role of the Pauli coupling of the vector meson
$\rho$ to the nucleon vis-a-vis the eventual softening of the EOS as
revealed through a substantial reduction of the incompressibility
parameter ($K_0$) for symmetric nuclear matter.
\end{abstract}

\newpage

The equation of state for nucleonic matter has been attracting considerable
attention from various perspectives ranging from the physics of the atomic
nucleus to the underlying mechanism of supernovae explosions \cite{swesty94} and the 
properties of the remnant neutron stars \cite{prakash97}, as also for the understanding of
data on heavy ion collisions. Out of the many relevant aspects, for the 
present purpose, we shall focus our attention on the compression modulus 
($K_0$) of nuclear matter. At an experimental level
this is inferred, at normal nuclear densities, from giant monopole (breathing
mode ) resonances for which rather precise data is available particularly
from the Sn isotopes \cite{blaizot80}. Again, at an observational stratum, this compressional
characteristic has important bearings on the matter, momentum and energy
flows in heavy ion collisions. Furthermore, these notions play an essential 
role in the keenly
desired strong core bounce following supernovae collapse, which in turn
provides an important input for the estimation of the maximum masses of 
neutron stars.

In the present letter we discuss the special effect of the Pauli coupling
of the vector meson (in particular $\rho$) to the nucleon, on the EOS, 
concentrating on its implications in so far as the incompressibility parameter
of nuclear matter is concerned. The natural setting for such an investigation
is, of course, the relativistic quantum field theoretic approach to the
nuclear many body problem, wherein the nucleons interact through the exchange
of virtual mesons within a framework in which Lorentz covariance, causality
and retarded propagation of particles is inherently incorporated, and the
approximations endemic to traditional non-relativistic nuclear physics 
automatically avoided.
Moreover, the usual argument, forming
the basis for the use of non-relativistic approximations, that the binding
energy of nucleons in nuclear matter is small as compared to their rest mass,
fails to hold once one realizes that this small value actually stems from an
extremely sensitive cancellation of the scalar and vector parts of the
nuclear potential. The need for a departure from the conventional 
non-relativistic potential models is further accentuated by the fact that in 
nuclear matter the nucleon mass is significantly reduced thereby rendering 
the lower components of the Dirac spinors too large to be neglected.

Accordingly the appropriate starting point is a model Lagrangian density of
nucleons interacting with the scalar ($\sigma$), vector ($\omega$ and $\rho$)
and pseudoscalar ($\pi$) mesons, namely

\begin{equation}
{\cal L}_{int} = g_\sigma {\bar N}\phi_\sigma N -
 g_{\alpha} [{\bar{N}} \gamma _\mu \tau^\alpha 
 N - i\frac{\kappa _\alpha}{2M}{\bar{N}}
         \sigma_{\mu\nu}\tau^\alpha N\partial ^\nu]
V^\mu_\alpha -  
 \frac{g_\pi}{2 M}{\bar{ N}}\gamma_\mu \gamma_5 N\partial^\mu \phi_\pi
\end{equation}                                                   
where $\phi_\sigma$ and $\phi_\pi$ are the $\sigma$ and $\pi$ fields
respectively and the generic vector field is denoted by
$V^{\alpha}_\mu$, $\alpha$ running
from 0 to 3, indexes quantities relevant for
$\omega$ (when $\alpha = 0$) and for $\rho$ (with $\alpha$=1 to 3);
also $\tau_0=1$ and $\tau_i$ are the
isospin Pauli matrices. Here $\kappa_\alpha$ is the strength of the
Pauli coupling of the vector mesons to the nucleon field N. In the
context of the given mileau the non-interacting nucleon propagator in
the presence of a nuclear medium  at zero temperature with Fermi momentum
$k_F$, is given by
$G^0(k) = G^0_F(k)  + G^0_D(k)$,
wherein the first term
is the free propagator of a spin $\frac{1}{2}$ particle, 
while the second term
\begin{equation}
G^0_D(k) = (k\!\!\!/ + M) [\frac{i\pi}{E(k)}
\nonumber\\ \delta (k_0-E(k))\theta(k_F-|\vec k |)]
\end{equation}
which involve $\theta(k_F-|\vec k|)$ guarantees Pauli blocking.
Here
$E (\vert \vec k \vert) = \sqrt {\vert \vec k \vert^2 + M^2}$. Next,
interactions are taken into account by working out the dressed or in-medium
nucleon propagator with mesonic degrees of freedom in play, through the Dyson equation, namely
$G(k) = G^0(k) + G^0(k)\Sigma (k) G(k)$
where $\Sigma$, the proper self-energy, has the general structure given by
\be
\Sigma(k) = \Sigma^s(\k,k_0)-\gamma^0 \Sigma^0(\k,k_0)+\vec\gamma \cdot \vec k \Sigma^v(\k,k_0)
\ee
omitting a possible tensor piece $\Sigma^t$ which does not appear
as far as we are concerned here; also we write $\Sigma^\mu=(\Sigma^0, 
\vec k \Sigma^v)$. 
Solution of the Dyson's equation is given by
\be
G^{-1}(k) = \gamma_\mu (k^\mu + \Sigma^\mu(k)) -(M+\Sigma^s(k))
\ee
This leads to a modification of the effective nucleon
mass, momentum and energy of the nucleon in nuclear matter
$M_k^\ast = M + \Sigma^s(\k,E_k)$, $|{\vec k}^\ast| = |{\vec k}|(1+
\Sigma^v(\k,E_k))$ and $E_k^\ast=(|{\vec k}^\ast|^2 + M_k^{\ast 2})^\frac{1}{2}$
respectively.

The $\Sigma^s$, $\Sigma^0$ and $\Sigma^v$ components of the nucleon
self energy receive contributions from the mesonic interactions being
considered and we give below the terms arising from the $\rho$-meson as
the rest are well documented{\footnote{
It is necessary, however, to point out that for the contribution of the
pion one usually retains only the long range part omitting the contact
interaction which is a constant in momentum space.}} elsewhere
\bea
\Sigma^s_ \rho(\k, E_k) = \frac{(2 \lambda - 1)g_\rho^2}{8 \pi^2 \k} \int d\q\q \{ -2
\frac{\mqs}{\eqs}
\Theta (\k,\q) + \nonumber\\ 
3(\frac{\kappa}{2M})^2\frac{\mqs}{\eqs}
(2\k\q
-\frac{1}{2} m_\rho^2\Theta (\k,\q) )\nonumber\\ 
+3(\frac{\kappa}{2M}) (
\q\frac{\qs}{\eqs}\Theta (\k,\q) -2 \k \frac{\qs}{\eqs}\Phi (\k,\q) )\}
\eea
\bea
\Sigma^0_ \rho(\k, E_k) = - \frac{(2 - \lambda)g_ \rho^2}{4 \pi^2 m_ \rho^2} \int d\q\q^2- \frac{(2 \lambda - 1)g_\rho^2}
{8 \pi^2 \k} \int d\q \q \{\Theta (\k, \q) \nonumber\\
+(\frac{\kappa}{2M})^2(2\k\q 
-\frac{1}{2} m_\rho^2\Theta (\k, \q) )\}
\eea
\bea 
\Sigma^v_p(\k, E_k) =
- \frac{(2 \lambda - 1)g_\rho^2}{8 \pi^2 \k^2} \int d\q \q \{ 2 \frac{\qs}{\eqs}\Phi (\k, \q)+ 
\nonumber\\ 2 (\frac{\kappa}{2M})^2 ( \k\q\frac{\qs}{\eqs}\Theta (\k, \q) \nonumber\\ 
-\frac{\qs}{\eqs} (\k ^2 + \q ^2 -\frac{1}{2} m_\rho^2 )\Phi (\k, \q)
\nonumber\\ -3 (\frac{\kappa}{2M}) 
(\k \frac{\mqs}{\eqs} \Theta (\k, \q) - 2\q
\frac{\mqs}{\eqs}\Phi (\k, \q) )\}
\eea
where the form of the functions $\Theta (\k, \q)$ 
and $\Phi (\k, \q)$ are as given in ref. \cite{serot83}.
The isospin degeneracy factor $\lambda$ takes values $1$ and $2$ for 
neutron matter and nuclear matter respectively. Eqs.[4-6], augmented by the 
contributions from the other mesons ($\omega,\sigma, \pi $ etc.),
would constitute, in fact, a set of three coupled non-linear integral
equations since the integrands themselves depend on the self-energy
components through the factors $|{\vec q}^\ast|$, $E_q^\ast$, $M_q^\ast$
as well as the retardation terms $(E_k-E_q)^2$ arising from the meson
propagators. Furthermore, this system must be considered in conjunction
with the transcendental equation
$E_k=[\eks-\Sigma^0(k)]_{k_0=E_k}$
for the single particle energy spectrum. From the foregoing contributions
the energy density is easily calculable. Thus for example referring to refs
\cite{serot83}-\cite{serot82} for the other terms, the $\rho$-meson contributes,
\bea
{\cal E_ \rho} =- \frac{1}{8}(2- \lambda) 
\frac{g_\rho^2}{m_ \rho^2} \rho_B^2  - \frac{2 g_\rho^2(5 \lambda-4)}{(2\pi)^6}\int\frac{d^3 k}{\eks}\int
\frac{d^3 q}{\eqs} D_\rho(k-q) \nonumber\\ 
\{\frac{1}{2}-(E_k-E_q)^2 D_\rho(k-q) \} \{(2 \mks \mqs-
q^* \cdot k^*) \nonumber\\
+3 (\frac{\kappa}{2M})((k-q) \cdot k^* \mqs - (k-q) \cdot q^* \mks) \nonumber\\
+ \frac{1}{2}(\frac{\kappa}{2M})^2((k-q)^2 q^* \cdot k^* 
- 4 ((k-q) \cdot k^*)
((k-q) \cdot q^*) \nonumber\\
+3 (k-q)^2 \mks \mqs)\} \Theta (k_F-\q) \Theta (k_F-\k)
\eea
The above formulae define the Relativistic Hartree Fock theory in general
and introduces the rho mesons into the general framework.

It is perhaps appropriate at this point to state that the model with which
we are working was first proposed as a mean field theory (MFT) with a minimun
inventory of the isoscalar $\sigma$ and $\omega$ mesons within its ambit,
wherein the meson field operators were replaced by their vacuum expectation
values, namely $\langle \sigma\rangle \neq 0$ and $\langle \omega_0 \rangle
\neq 0$ ( as $\langle \vec \omega \rangle = 0$ in view of the isotropy of 
space ) \cite{walecka86}. This held forth the promise of providing a reasonable EOS at normal
densities which is easily extended to high density matter without introducing
intractable complications. In MFT the effective nucleon mass is not momentum
dependent, the nucleon momentum remains unmodified and the single particle
spectrum is readily determined. The pion field is unable to appear in
MFT, because it is a pseudoscalar, and nuclear matter is parity invariant. 
As for the $\rho$ meson $\langle \vec\rho\rangle = 0$,
while $\langle \rho_0 \rangle \neq 0$ only if the underlying nuclear mater
is isospin asymmetric (viz. the neutron and proton densities are unequal) \cite{toki94}.

We are now in a position to focus our attention on the incompressibility
parameter $(K_0)$ vis-a-vis the Pauli coupling ($\kappa_\rho$) of the 
rho-meson to the nucleon. Concentrating on symmetric nuclear matter our
strategy shall be to vary the parameters $g_\sigma$ and $g_\omega$ to fit
the saturation density $\rho_0$ ( corresponding to $k_F = 1.4 fm^{-1}$) and the
binding energy per nucleon $\epsilon$ (15.75 MeV), and then to compare the
modulus of compression $(K_0)$ in different scenarios, in order to understand
the role of the Pauli coupling of the vector meson, and this is displayed
in Table-I.
\vskip 0.2 cm 
\centerline{Table-I : Coupling parameters and incompressibility}
\par
\begin{tabular}{l|r|r|r|r|r|r|r|r}
\cline{2-7}
\verb| |
&Models&$g_\sigma$& $g_\omega$ & $g_\rho$ & $\kappa_\rho$ & $K_0$ (MeV)\\
\cline{2-7}
&MFT & 9.583 & 11.685 & -- & -- & 550.00 \\
&HF & 9.126 & 10.405 & 0.0 & 0.0 & 598.01 \\
&HF & 8.846 & 10.203 & 2.63 & 0.0 & 569.83 \\
&HF & 9.196 & 10.048 & 2.63 & 3.7 & 355.12 \\
\cline{2-7}
\end{tabular}
\par

One observes right away that rather little adjustment in $g_\sigma$ and 
$g_\omega$ are necessary to fit $\rho_0$ and $\epsilon$ in diverse situations.
This again is due to the sensitive cancellation of those two contributions.

The first row depicts the results of the original MFT where, of course, in
the case of symmetric nuclear matter, the iso-vector $\rho$ is unable to
contribute. The next set of entries pertains to the inclusion of the
exchange terms (HF) with only $\sigma$ and $\omega$ allowed to enter the
fray. Here one is invited to observe the increase in the parameter $K_0$,
indicating the exchange terms being repulsive, leads to a hardening of the
EOS. This is also reflected in the greater reduction of the repulsive 
$\omega$ as compared to the attractive $\sigma$. In the third row of the
table we introduce the isovector $\rho$ meson ( but without the Pauli 
coupling) and observe the reduction of $g_\sigma$ indicating the attractive
nature of the minimal $\rho$ exchange ( in the context of symmetric nuclear
matter ), which naturally leads to a softening ( as apparent from the 
corresponding $K_0$ value) with the introduction of the Pauli 
coupling($\kappa_\rho$) as shown in the last row of the table.
It is clear that a
further source of attraction has appeared and the value of the
parameter $K_0$ has been dramatically reduced, thus leading to the 
conclusion that the Pauli coupling brings about an overall softening of the 
EOS. 

We have kept the value of $\kappa_\rho$ fixed at 3.7 ( as suggested by vector
meson dominance picture of Sakurai) to fit the anamalous iso-vector magnetic
moment of the nucleon \cite{bouyssy87}. If one were to use the value recommended by the 
Bonn group the softening would be even more dramatic. 

Here it may also be mentioned that the pions are found to provide
more binding provided the zero range interactions emanating therefrom 
are excluded. The main conclusions discussed above are also pictorially
discernable from Fig.1. One notes how the binding energy per nucleon
and the saturation density have been pinned down to their observed values in the
different scenarios, and how the inclusion of the Pauli coupling has led 
to a shallower minimum, a decrease in the curvature of the energy 
against the Fermi momentum.

In conclusion, it may be noted that the Pauli coupling can only contribute
provided one includes exchange terms which in turn gives rise to the 
further momentum dependence in the self energies and plays an appreciable
role in softening the E.O.S. Even though, as we have
seen, $\Sigma_s$ , $\Sigma_0$ and $\Sigma_v$ have only weak momentum
dependence, it has a significant role so far as the EOS of nuclear matter
is concerned as the binding energy again emanates from a very sensitive
cancellation of scalar and vector self energies. Further implications of
the exchange correction and the role of the Pauli coupling on the general
EOS of asymmetric nuclear matter and different regimes of 
nuclear matter densities shall be presented elsewhere.

\newpage
\noindent
{\bf Figure Caption:}
\begin{enumerate}

\item Binding energy vs Fermi momenta ($k_F$ fm$^{-1}$) curve with
      solid line representing MFT result while the
dotted, dashed, and dashed dotted lines depict Hartree Fock calculations 
      HF($\sigma, \omega$), HF1($\sigma, \omega, \rho$),
      and HF2($\sigma, \omega, \rho$) respectively.
      HF1($\sigma, \omega, \rho$) displays results without 
Pauli coupling while HF2($\sigma, \omega, \rho$) potrays results with
Pauli coupling.
\end{enumerate}
   
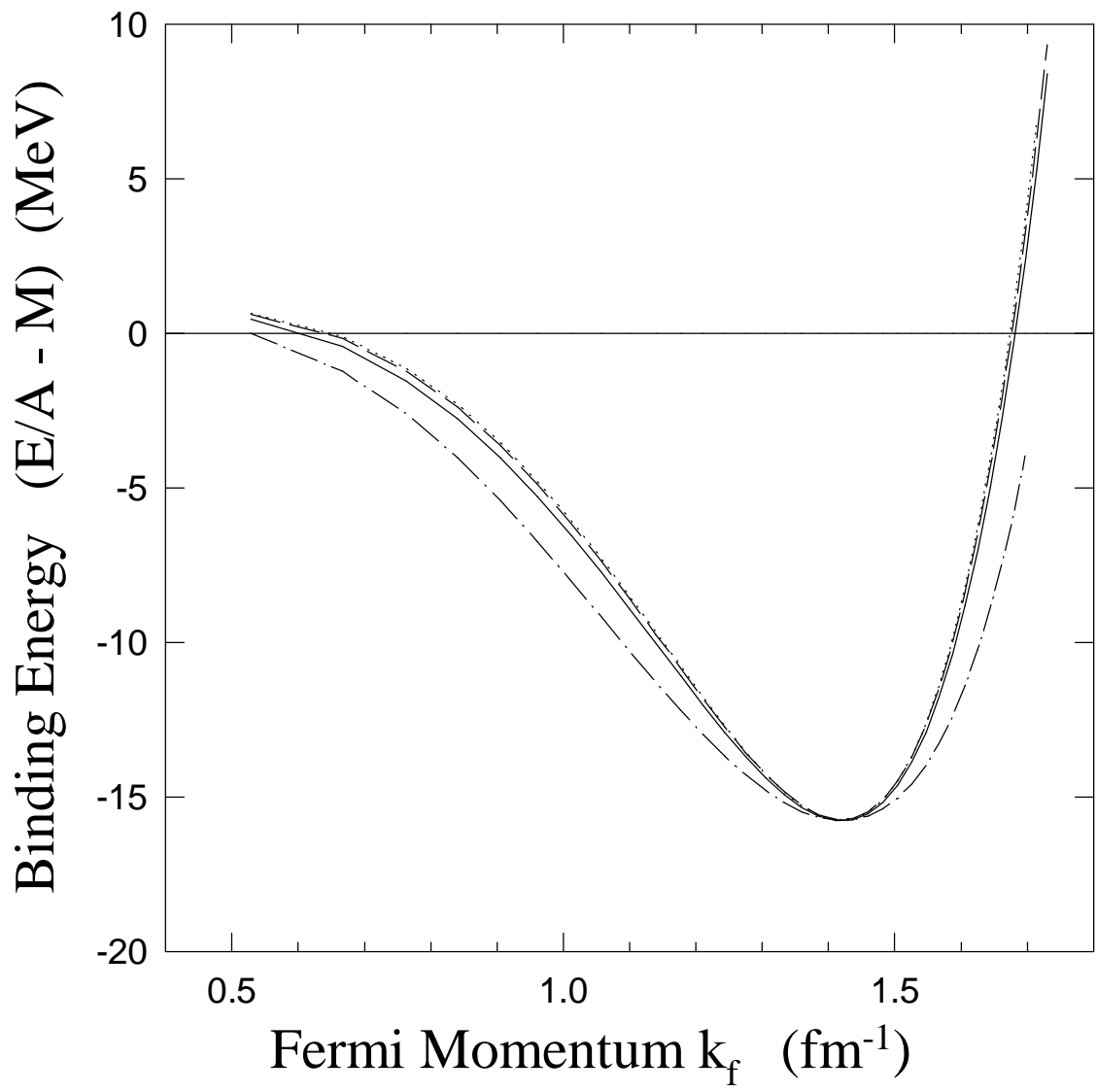
\end{document}